\documentclass{appolb}
\usepackage{epsfig}

\date{\today}
\newcommand{\be}{\begin{equation}}
\newcommand{\ee}{\end{equation}}
\newcommand{\bc}{\begin{center}}
\newcommand{\ec}{\end{center}}
\newcommand{\ba}{\begin{array}}
\newcommand{\ea}{\end{array}}

\begin{document}
\title{CRANKING IN ISOSPACE - TOWARDS A CONSISTENT MEAN-FIELD DESCRIPTION
 OF $N=Z$ NUCLEI\thanks{Invited talk presented at the International Conference
{\it High Spin 2001\/}, Feb. 6-11, 2001, Warszawa, Poland}
}
\author{{\sc Wojciech Satu{\l}a and {\sc Ramon A. Wyss}}
\address{Institute of Theoretical Physics, University of Warsaw,\\
             ul. Ho{\.z}a 69, PL-00 681 Warsaw, Poland \\
         Royal Institute of Technology,
             Physics Department Frescati,\\
             Frescativ{\"a}gen 24, S-104 05 Stockholm, Sweden}
}
\maketitle

\begin{abstract}

Excitation spectra $\Delta E_T$ of $T=0,1,2$ states in even-even (e-e) and 
odd-odd (o-o) $N=Z$ nuclei are analyzed within a mean-field based model 
involving isovector and isoscalar pairing interactions and 
the iso-cranking formalism
applied to restore approximately isospin symmetry. 
It is shown that $T=0$ states in o-o  and $T=1$ states in e-e
nuclei correspond to two-quasiparticle,
time-reversal symmetry breaking excitations  since 
their angular momenta are $I\ne 0$. On the other hand 
the lowest $T=2$ states in e-e and $T=1$ states in o-o nuclei,
which both are similar in  structure   to their 
even-even isobaric analogue states, are described as
e-e type vaccua excited (iso-cranked) in isospace.
 It appears that in all cases isoscalar 
pairing plays a crucial role in restoring the proper value of the inertia
parameter in isospace i.e. $\Delta E_T$.

\end{abstract}


\section{Introduction}

The theoretical treatment of the generalized pairing problem 
is interesting and clearly nontrivial. Though its fundaments 
in the form of generalized BCS (or HFB) theory were laid down almost 
thirty years ago in a number of papers by different 
groups~\cite{[pn60]} there
are still many open problems. 
They cover a broad range of questions starting from the form of 
{\it effective\/} $pn$-pairing 
interaction and 
structure of {\it effective\/} $pn$-Cooper pairs,
 to problems related to symmetries 
and symmetry restoration, 
problems of interplay between 
the quasiparticle and isospin degrees of freedom to the role played by    
higher-order effects like
quartetting or $\alpha$-type condensation.
It includes also the fundamental question
concerning the experimental fingerprints of $pn$-collectivity.

Thus far, the strongest indications of $pn$ (isoscalar) 
collectivity come from: ({\it i\/}) shifts of crossing frequencies
in gsb of some e-e $N=Z$ nuclei [the best studied case is $^{72}$Kr]
and ({\it ii\/})  the mass defect in $N\sim Z$ nuclei commonly known 
as the Wigner energy problem. 
The latter effect can be locally
restored by enforcing collective isoscalar $pn$-pairing~\cite{[Sat97]}. 
Based on such a model 
we aim here to look into the impact of collective $pn$ correlations
upon the structure of $T=0,1,2$ excitations in $N=Z$ nuclei. 
The goal is to reveal the interplay
between quasiparticle and isospin excitations
in $N=Z$ nuclei and role played by isoscalar $pn$-collectivity. 
The present paper supplements our previous 
letters~\cite{[Sat00a],[Sat00b]}.

\section{The paradigm of the Wigner energy}

\begin{figure}
\begin{center}
\leavevmode
\epsfysize=10.0cm
\epsfbox{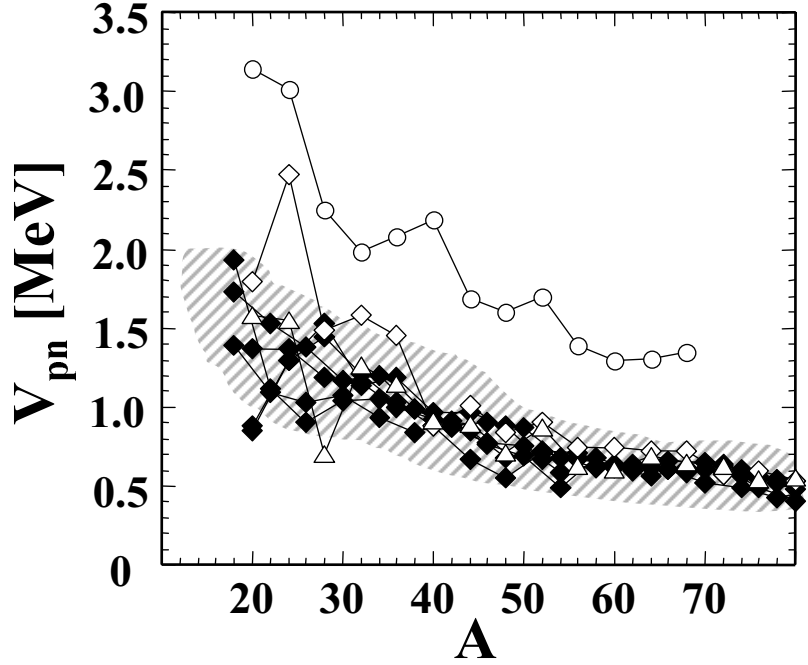}
\end{center}
\caption[]{Empirical and theoretical  values of
$V_{pn}\approx {\partial^2 B\over \partial N\partial Z}$
[for both 
spherical and deformed HFB calculations].
Solid symbols mark $V_{pn}(A)$ for e-e $N-Z=2,4$. In these 
cases $V_{pn}$ probes essentially the quadratic term in 
the nuclear symmetry energy $E_{sym} \sim (N-Z)^2$
and both empirical and theoretical curves do roughly 
overlay each other. For e-e $N=Z$ nuclei a strong 
enhancement of $V_{pn}$ 
is seen in the empirical data ($\circ$). This apparent Wigner 
energy effect is not at all seen in the spherical calculations 
($\triangle$) and only modestly visible in 
deformed calculations ($\diamond$).}
\label{hfwig}
\end{figure}

Traditional mass models based on the mean-field approach strongly
underbind $N\sim Z$ nuclei~\cite{[Mye66],[ETFSIs],[Sat97as]}.
This  additional binding energy which is known as the
Wigner energy, is usually parametrized as~\cite{[Jen84]}:
\begin{equation}\label{wigner}
E_W = W(A)|N-Z| + d(A)\delta_{NZ}\pi_{pn}   \quad
\mbox{where} \quad \pi_{pn}= \left\{ \begin{array}{ll}
1 & \mbox{for odd-odd nuclei} \\  0 & \mbox{otherwise} \end{array} \right.
\end{equation}
and $W(A) \sim 47/A$\,MeV~\cite{[Sat97as]}.
A  microscopic explanation of the Wigner energy within the
\underline{\it mean-field\/} approach is still lacking.
The 'congruence' energy mechanism proposed in~\cite{[Mye97]}
is in its nature a geometrical concept
[i.e. independent on a specific form of nuclear interaction].
There one assumes an  enhanced particle-hole ($ph$) interaction at the
$N\sim Z$ line which, however, is not at all present
in traditional \underline{{\it spherical\/}} Skyrme-HFB
calculations, see Fig.~\ref{hfwig}. This is basically due to the
pairing interaction which evenly distributes particles over spherical
subshells thus strongly averaging over the properties of
neighbouring nuclei.  One would expect
the effect to show up [and it indeed does]
in \underline{{\it deformed\/}} microscopic
calculations. However, as shown in Fig.~\ref{hfwig} for a particular set
of SIII+BCS mass calculations of Ref.~\cite{[Taj96s]} it is seen
only in light nuclei  and accounts for at most
20\% of the empirical Wigner energy strength $W(A)$.
Even in the limit of single-particle ({\it sp\/}) Skyrme-HF 
calculations one can
account for at most 30\% of the empirical value.

It seems therefore, that a microscopic scenario based on
isoscalar $pn$-pairing~\cite{[Sat97]} is the most promising
so far. It requires [within the mean-field model] the isoscalar pairing
to be on the average stronger than the isovector. Although there are clear
empirical~\cite{[Sat97as]} as well as
theoretical~\cite{[Bre90],[Sat97as],[Rop00]} arguments that the
Wigner energy is indeed due to the isoscalar interaction
it is not at all settled whether it is due to a static pairing effect.
In spite of that, in the following we will enforce the 
strength of the isoscalar pairing
interaction so as to reproduce the Wigner energy and will 
 look into the impact of these correlations
 on the structure of isobaric excitations in $N=Z$
 nuclei.

\section{The model}

Our model Hamiltonian contains the Woods-Saxon $sp$ potential and a schematic
pairing interaction including isovector ($t=1$) and
isoscalar \hbox{($t=0$)}
terms\footnote{Throughout the text small letter $t$ is reserved to label
the type of pairing correlations while capital $T$ 
corresponds to the total nuclear isospin.} of the form:
\be
H_{pair} = - G^{t=1} P^\dagger_{1\mu} P_{1\mu} - G^{t=0}
P^\dagger_{00} P_{00} 
\ee 
where $P^\dagger_{tt_z}$ create isovector and isoscalar
pairs in time-reversed orbits: 
\be\label{pairs} 
P^\dagger_{1\pm 1} =
a^\dagger_{\alpha\tau} \bar a^\dagger_{\alpha\tau};\quad P^\dagger_{10(00)} =
{1\over \sqrt2} (
               a^\dagger_{\alpha n} \bar a^\dagger_{\alpha p} \pm
               a^\dagger_{\alpha p} \bar a^\dagger_{\alpha n} ).
\ee
Note that our interaction does not ascribe any specific
structure [e.g. deuteron like] to the $pn$ pairs
but only counts their effective number. The
Bogoliubov transformation which is used [$\alpha$ runs over {\it sp\/} states,
$\tau$ denotes third component of isospin, and $k$ labels the quasiparticles]:
\begin{equation}\label{bogol} 
\alpha^\dagger_k =
\displaystyle\sum_{\alpha\tau >0} (U_{\alpha\tau , k} a^\dagger_{\alpha\tau}
+V_{\bar\alpha\tau , k} a_{\bar\alpha\tau }+ U_{\bar\alpha\tau ,  k}
a^\dagger_{\bar \alpha\tau }
+V_{\alpha\tau , k} a_{\alpha\tau})
 \end{equation}
 is the most general one,
allowing for an unconstrained mixing of neutron and proton holes and particles.
The problem is solved using the Lipkin-Nogami (LN)
approximate number projection technique. In this respect the present model
follows rather closely the description of Ref.~~\cite{[Sat00]}.
   The use of the LN model allows for mixing of both $t=1$ and $t=0$ phases
  at the cost of spontanous isospin symmetry breaking~\cite{[Sat97],[Sat00]}.
  Generally, the mechanism of spontanous symmetry breaking 
  is the only mechanism which allows
  to take into account correlations which are seemingly beyond
  the mean-field. Obviously, 
  the best example is  the spontanous breaking
  of spherical symmetry. Without it the {\it mean-field\/} would be capable
  to describe only a few nuclei.
  In our opinion the spontanous breaking of isospin symmetry brings
  our solution closer to reality and allows for simulation of higher order
  effects like quartetting or $\alpha-$clustering.
  These effects are present in exact-model solutions~\cite{[Dob98]}
  which always do mix $t=0$ and $t=1$ phases in contrast to the
  generalized BCS approximation~\cite{[Eng96],[Eng97],[She99]}. 
  Moreover, in the
  following we will approximately restore this symmetry by applying the
  isospin cranking mechanism 
to generate isospin $T=0,1,2$ states in $N=Z$ nuclei.

In the following calculations we will freeze the deformation degree
of freedom of our Woods-Saxon potential at $\beta_2 = 0.05$. We compute
$G^{t=1}(A)$ using the average gap method of Ref.~\cite{[Mol92]} and taking
a symmetric cut-off including the lowest $[A/2]$  neutron and
 $[A/2]$ proton $sp$ states.
The strength of isoscalar $G^{t=0}(A)$ pairing is computed by means of
a direct fit to the Wigner energy strength $W(A)\sim 47/A$\,MeV, 
see~\cite{[Sat00a]} for details. This method assumes that $W(A)$ is
entirely due to the isoscalar pairing field. Since we use here an 
almost spherical
mean-field this assumption seems to be consistent with the conclusions of
the preceeding section. Deformation effects are expected to result in 
a rather modest reduction of $G^{t=0}$.
 Let us mention that the same procedures and parameters are
 used systematically in all presented calculations and that they exactly
 follow Refs.~\cite{[Sat00a],[Sat00b]}.

\section{Isobaric excitations in $N=Z$ nuclei}

The isoscalar pairing field naturally takes into account the mass-excess in
$N=Z$ nuclei. The aim of this section is to show that it is also of vital
importance to understand  the excitation energy pattern, $\Delta E_T$, of
the elementary isospin  $T=0,1,2$ excitations in even-even (e-e) and
odd-odd (o-o) $N=Z$ nuclei. To compute these excitations,
which follow a $\Delta E_T\sim T(T+1)$ pattern, we invoke the technique
of cranking
in isospace in analogy to spatial rotations. The discussion is organized as
follows: ({\it i\/}) the pure single-particle model will be succeeded
by ({\it ii\/}) the standard, isovector $nn-$ and $pp-$ paired model, and
eventually ({\it iii\/}) by the isovector and isoscalar paired model.

\subsection{The extreme single-particle model}\label{spmodel}

Let us consider the isospin-cranked $sp$ model:
$\hat H^{\omega_\tau} =
\hat H_{sp} -\hbar \omega_\tau \hat t_x$. Let us further assume for
simplicity that $\hat H_{sp}$ generates a fixed, equidistant spectrum
of 4-fold degenerate levels [including isospin and Kramers degeneracy]
$e_i = \delta e i$. The isospin-cranking removes the isospin
degeneracy. The quartets of $sp$-states
split into two pairs of Kramers degenerate routhians
$|\pm\rangle = {1\over \sqrt2}(|n\rangle \pm |p\rangle$.
The  slope of the 
pair of $|+\rangle$ ($|-\rangle$) routhians is determined
by their $sp$ alignment in iso-space
 $\langle\pm | \hat t_x | \pm \rangle = \pm 1/2$, respectively.

\begin{figure}
\begin{center}
\leavevmode
\epsfysize=10.0cm
\epsfbox{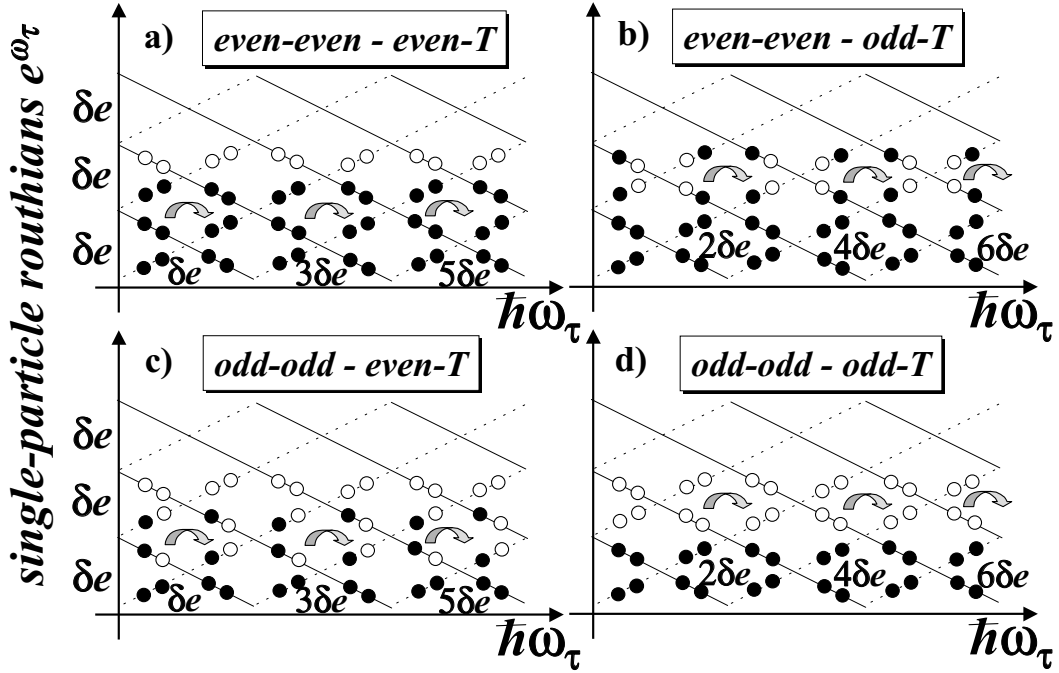}
\end{center}
\caption[]{Single-particle routhians representative for the following cases:
({\bf a}) e-e nucleus
even-T, ({\bf b}) e-e nucleus odd-T case, ({\bf c})
o-o nucleus even-T case, and ({\bf d})  o-o nucleus odd-T case.
Filled circles mark occupied states. Arrows indicate configuration
changes versus iso-frequency.
}
\label{spmod}
\end{figure}

Fig.~\ref{spmod} shows the $sp$ routhians for the lowest $sp$ configurations
in even-even (upper part) and odd-odd (lower part) $N=Z$ nuclei.
For the case of the even-even vacuum [Fig.~\ref{spmod}a] one obtains
configuration changes at iso-frequencies:
$\omega_\tau^{(c)}=\delta e,3\delta e,...,(2n-1)\delta e$. At
each crossing a pair of upsloping routhians become empty
and a pair of downsloping routhians become occupied. This reoccupation
process  gives rise
to stepwise changes in iso-alignment in units of $\Delta T_x$=2.
Since simultanously $T_y=T_z=0$, one obatins a sequence
of even-isospin states $T=0,2,4,...,2n$. The odd-T sequence
of states in even-even
$N=Z$ nuclei can therefore be reached only for excited states. The
lowest particle-hole excitation leading to odd-T states is shown
in Fig.~\ref{spmod}b. Indeed, the initial iso-alignment
of this state is $T_x=1$ and, as in the case of the e-e vacuum discussed
above, it is changed in steps
of $\Delta T_x$=2 but at iso-frequencies
$\omega_\tau^{(c)}=2\delta e,4\delta e,...,2n\delta e$.
Simple calculations give the following expression for
the total energy
\begin{equation}
E \equiv E^\omega + \omega_\tau T_x = {1\over 4} \delta e [1-(-1)^{T_x}]
+ {1\over 2} \delta e T_x^2.
\end{equation}
Both odd and even-T bands have therefore the same
inertia parameter [reciprocal of the moment of inertia $a=\Im^{-1}$]
$a =\delta e$ which is proportional to the average level spacing
at the Fermi energy. The bands are  shifted by
$\Delta E = E_{T=1} - E_{T=0} = \delta e$ i.e. by a {\it ph\/}
excitation energy.

For the case of o-o nuclei the last quartet is only half-filled. Hence,
two different {\it sp-}configurations can be obtained: ({\it i\/})
an aligned one of $T_x$=1 [Fig.~\ref{spmod}d]
and ({\it ii\/}) a
non-aligned one of $T_x$=0 [Fig.~\ref{spmod}c].
Since the reoccupation taking place at high iso-frequency
always gives rise to a stepwise change in total isospin
in units of $\Delta T_x$=2 the aligned configuration
gives rise to an odd-T sequence of states while the non-aligned 
one builts up
the even-T sequence of states. 
Observing further that crossings take place
at $\hbar\omega_\tau^{(c)}= 2n\delta e$ [case ({\it i\/})]
and $\hbar\omega_\tau^{(c)}= (2n-1)\delta e$ [case({\it ii\/})] it is
strightforward to compute the total energy:
\begin{equation}
E  = -{1\over 4} \delta e [1-(-1)^{T_x}]
+ {1\over 2} \delta e T_x^2.
\end{equation}
Two very interesting conclusions arise from this discussion. First of all,
the {\it docoupling effect\/} is so strong in this case
that it gives rise to a complete degeneracy of $T$=0 and $T$=1 states.
This is in nice qualitative agreement with the data since $T$=0 and $T$=1
states are indeed nearly degenerate  in o-o
but not in e-e nuclei~\cite{[Jan65],[Zel76]}. Secondly,
let us observe that the aligned configuration {\it does not break time-reversal
invariance\/} while the non-aligned configuration {\it does break it\/}.
Again this is in agreement with
empirical data since all $T=0$ states have non-zero
angular momentum $I\ne0$ while $T=1$ states have $I=0$. 
In other words the theoretical treatment of $T=0$ states 
requires the explicit breaking of time-reversal invariance.
In the presence of pairing correlations it means that $T=0$ states
should be treated as two-quasiparticle
({\it 2qp\/}) excitations. Treating them on the same footing
as the neighbouring e-e vaccua would correspond to what is
usually known in the literature as the {\it filling approximation\/}.
Within the {\it filling approximation\/} pairs of 
{\boldmath{$\alpha\alpha$}} type
are always accompanied [with the same occupation probability] by pairs 
of {\boldmath{$\bar\alpha\bar\alpha$}} type forming a time-reversal invariant
many-body state.

Note also, that the same situation applies for the
$T=1$ states in e-e nuclei. These states, within the {\it sp-}model
are {\it ph\/} excitations. Therefore, they do breake time-reversal
invariance and, in the presence of pairing correlations, correspond to
{\it 2qp\/} configurations.

\subsection{Influence of T=1 pairing}

Let us investigate next the influence of standard $nn$ and $pp$ isovector
pairing correlations on the $sp$ iso-alignment processes and
iso-inertias discussed in the preceeding subsection. Let us still
consider the equidistant level model and assume
that the pairing gaps $\Delta_{pp}=\Delta_{nn}=\Delta$ are constant
as a function of $\hbar\omega_\tau$.
In the gap-non-self-consistent regime the
BCS equations can be solved analytically. The eigenenergies 
(positive) are [$\tilde e_\alpha \equiv e_\alpha -\lambda$]:
\be
E_{\alpha,\pm} [\equiv E_{\bar\alpha,\pm}]
= \sqrt{ [\tilde e_\alpha \pm {1\over 2}\hbar\omega_\tau ]^2
  + \Delta^2 },
\ee
and the associated eigenvectors:
\be\label{ev}
   \left[ \ba{c} U_{\alpha,n} \\ U_{\alpha,p}  \\
                 U_{\bar\alpha,n} \\ U_{\bar\alpha,p} \\
                 V_{\alpha,n} \\ V_{\alpha,p}  \\
                 V_{\bar\alpha,n} \\ V_{\bar\alpha,p}
          \ea \right]: \quad
  \longrightarrow \quad
   \left[ \ba{c} U^{(+)}_\alpha \\  -U^{(+)}_\alpha  \\ 0 \\ 0 \\
          0 \\ 0 \\ V^{(+)}_\alpha \\ -V^{(+)}_\alpha \ea  \right] ,
   \left[ \ba{c}  0 \\ 0 \\  U^{(+)}_\alpha \\  -U^{(+)}_\alpha \\
          V^{(+)}_\alpha \\  -V^{(+)}_\alpha \\ 0 \\ 0 \ea \right] ,
   \left[ \ba{c} U^{(-)}_\alpha \\   U^{(-)}_\alpha  \\ 0 \\ 0  \\
          0 \\ 0 \\ V^{(-)}_\alpha \\ V^{(-)}_\alpha \ea   \right] ,
   \left[ \ba{c}  0 \\ 0 \\  U^{(-)}_\alpha \\  U^{(-)}_\alpha  \\
          V^{(-)}_\alpha \\ V^{(-)}_\alpha \\ 0 \\ 0 \ea   \right] ,
\ee
where:
\be
  U^{(\pm)}_\alpha = {1\over 2}\sqrt{ 1 +
   {\tilde e_\alpha \pm {1\over 2}\hbar\omega_\tau \over E_{\alpha,\pm} }
   } \quad \mbox{and} \quad
  V^{(\pm)}_\alpha = {1\over 2}\sqrt{ 1 -
   {\tilde e_\alpha \pm {1\over 2}\hbar\omega_\tau \over E_{\alpha,\pm} }
   }
\ee
   It is interesting to observe that the solutions of the {\it sp\/} model
   $|\alpha;\pm\rangle$ form, in this case, the canonical basis. Indeed,
   see (\ref{ev}), the quasiparticle operators take the following
   structure:
\be
\alpha^\dagger_{\alpha,\pm} = \sqrt2 U^{(\pm)}  a^\dagger_{\alpha ,\mp}
          +  \sqrt2 V^{(\pm)}\bar a_{\alpha ,\mp} \quad
          \mbox{where} \quad   a^\dagger_{\alpha ,\pm}
          ={1\over \sqrt2} ( a^\dagger_{\alpha ,n}
          \pm   a^\dagger_{\alpha ,p} ).
\ee
  Let us further observe that this solution does also preserve
 iso-signature [$R_\tau = \exp^{-i\pi t_x}$] as a 
  self-consistent symmetry.

\begin{figure}
\begin{center}
\leavevmode
\epsfysize=10.0cm
\epsfbox{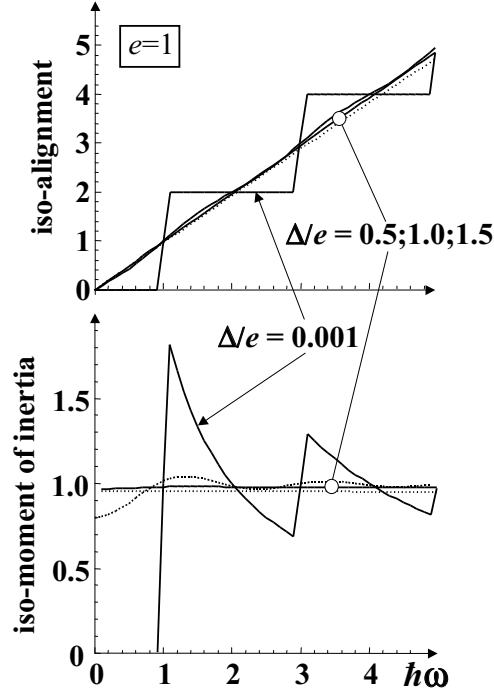}
\end{center}
\caption[]{ Iso-alignment ({\bf a}) and iso-MoI
({\bf b}) calculated using Eq.~(\ref{inert}) for very weak
 $\Delta= 0.001\delta e $, weak $\Delta = 0.5\delta e $,
 intermediate $\Delta = \delta e $  and strong 
$\Delta = 1.5\delta e$ pairing as
 a function of iso-frequency.
}
\label{pair}
\end{figure}

  It is
  relatively simple to derive an analytical expression for  
  the kinematical iso-moment of
  inertia  $\Im = T_x/\omega_\tau$. In case of the symmetric basis cut-off
  it reads:
\be\label{inert}
\Im = {1\over 2\omega_\tau}
\displaystyle\sum_{i=1}^{A/4} \left\{ {(2i-1)\delta e + \hbar\omega_\tau  \over
 \sqrt{ {1\over 4}[(2i-1)\delta e +\hbar\omega_\tau]^2  + \Delta^2 } } -
 {(2i-1)\delta e - \hbar\omega_\tau  \over
 \sqrt{ {1\over 4}[(2i-1)\delta e -\hbar\omega_\tau]^2  + \Delta^2 } }
                            \right\}
\ee
  In the low-frequency limit one obtains:
\be
 \Im \sim
\displaystyle\sum_{i=1}^{A/4} { \Delta^2  \over
 \left[ {1\over 4}[(2i-1)\delta e ]^2  + \Delta^2  \right]^{3/2} }
   \sim {1\over \delta e}
 \left\{
  \ba{rcc} 0.80 & \quad \mbox{for} \quad & \Delta={1\over 2}\delta e  \\
           0.96 & \quad \mbox{for} \quad & \Delta=          \delta e  \\
           0.90 & \quad \mbox{for} \quad & \Delta=         2\delta e  \ea
                                                               \right.
\ee
i.e. $\Im \sim \Im_{sp}$ almost independently [within reasonable range] on the
magnitude of pairing correlations, see also Fig.~\ref{pair} 
showing the numerical results
of the exact formula (\ref{inert}). Taking into account the 
self-consistency of the
pair gap does not change this conclusion as it was demonstrated within
the LN approximation in~\cite{[Sat00a]}.

The isovector {\it nn-} and {\it pp-}pairing introduces
a kind of {\it collectivity\/} on top of the discrete {\it sp\/} solutions.
In order to take into account quantum fluctuations in a similar way
as for spatial rotations, the cranking constraint in our
calculations corresponds to
$T_x = \langle \hat t_x\rangle = \sqrt{T(T+1)}$.

\begin{figure}
\begin{center}
\leavevmode
\epsfysize=10.0cm
\epsfbox{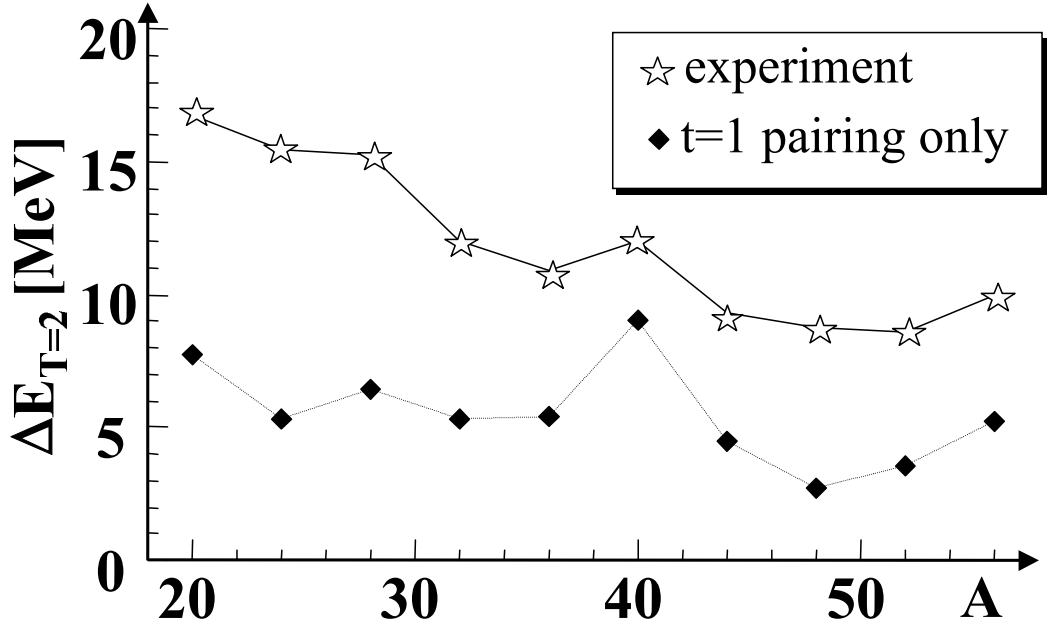}
\end{center}
\caption[]{ Empirical (stars) and calculated (filled circles)
excitation energies of $T=2$ states in e-e $N=Z$. These calculations
do include only $t=1$ pairing correlations which give rise
to large discrepancies
between theory and experiment.
}
\label{t2t1}
\end{figure}

Following the results of the $sp$ model
we have calculated $T=2$ states in e-e nuclei by means of
iso-cranking the quasiparticle vacuum to
$T_x = \sqrt6$. The results of the calculations
are shown in Fig.~\ref{t2t1}. In these calculations we used the 
static $sp$ spectrum
of the weakly [$\beta_2 = 0.05$] deformed Woods-Saxon potential plus standard
isovector $pp-$ and $nn-$pairing interaction. The results of the
calculations are far below the empirical data. It is rather obvious
that this level of disagreement must be related to a
physics mechanisms which is beyond our model.
First of all the {\it ph\/}
isovector terms are missing. At present we do not have any control  over
these effects. However it is very unlikely that they can
fully cure the problem since the
self-consistent mean-field models which do include the isovector {\it ph\/}
channel yield:
\be
E_{sym}(A,T) \approx {1\over 2} a_{sym} {T^2\over A} \equiv
{1\over 2} \left[ {T^2\over T(T+1)} a_{sym} \right] {T(T+1)\over A}
\ee
i.e. provide only ${T^2\over T(T+1)}$ fraction of the
empirical symmetry energy\footnote{In fact,
the empirical data in $N\sim Z$ nuclei are consistent with T(T+$\lambda$)
dependence for the symmetry energy
 where $\lambda \sim 1.25$~\cite{[Jan65]}.}

\subsection{The influence of t=0 pairing}

The condensate
of {\it pn\/} Cooper pairs with isospins coupled
antiparallel  is expected to
lower considerably the iso-moment-of-inertia (iso-MoI) in a similar
fashion as {\it nuclear superfluidity\/} influences the spatial MoI.
Furthermore, in response to iso-rotations, the isoscalar paired nucleus
is expected to undergo a phase transition due to the iso-CAP (Coriolis
Anti-Pairing effect) similar to the standard CAP effect which is so well
documented in high-spin physics.
The situation here is  probably even simpler
as compared to the case of spatial rotations.
There, the phase transition is always heavily modified due to the  
strong dependence of CAP on the orbital
 angular momentum [Stephens-Simon effect~\cite{[Ste72]}].
 In isospace one would expect a phase transition to be
  of bulk type resembling much closer the
  Meissner effect known in macroscopic superconductors~\cite{[Mei33]}.

 \subsubsection{The T=2 states in even-even nuclei}\label{st2}

\begin{figure}
\begin{center}
\leavevmode
\epsfysize=10.0cm
\epsfbox{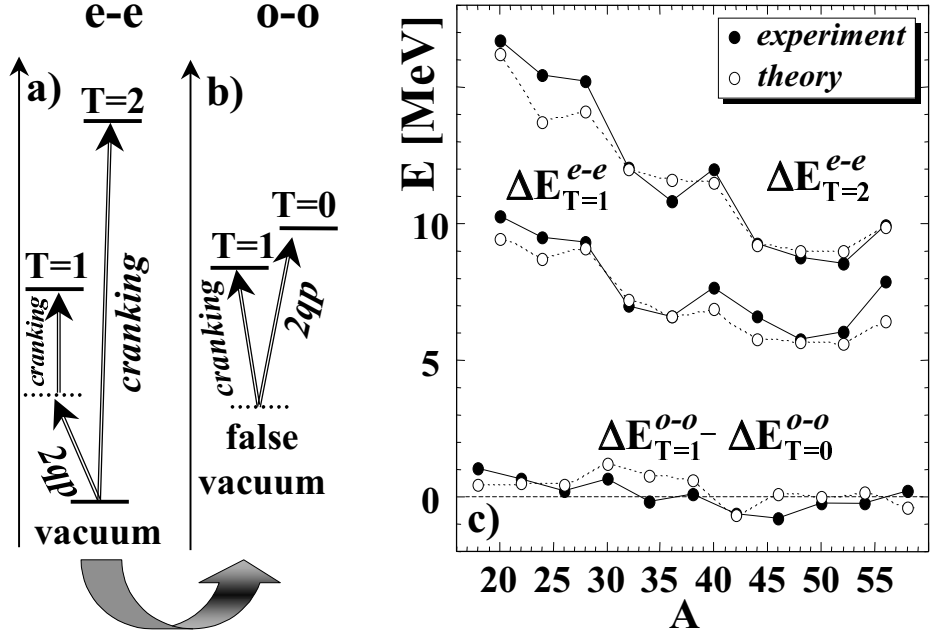}
\end{center}
\caption[]{
Schematic illustration of the calculation scheme applied to compute
$T=1,2$ states in e-e $N=Z$ nuclei {\bf a)} and $T=0,1$ states in o-o
$N=Z$ nuclei {\bf b)}. Part {\bf c)} shows the calculated
and experimental excitation energies of these states.
}
\label{all}
\end{figure}

 To verify these ideas we have performed a set of microscopic
 calculations for $T=2$ states in $8<N=Z<30$ e-e nuclei using
 the LN model which includes both isovector and isoscalar
 pairing correlations. The $T=2$ states  were computed by
 iso-cranking the e-e vacuum to iso-spin
 $T_x=\sqrt6$, see Fig.~\ref{all}a.
 These calculations nicely reveal all  features of $t=0$ pairing expected
 from our intuitive considerations. The  iso-MoI are indeed considerably
 lowered by $t=0$ pairing. At large iso-frequencies, but systematically
 below
 $T_x = \sqrt6$, we observe a phase transition from the $t=1$ and $t=0$ paired
 system to the $t=1$ paired system. The transition is indeed sharp, indicating
 its bulk character although one can always argue that fluctuations
 can smear it out, as it is usually the case in finite many-body system.
 In spite of that one can rather safely state that $T=2$ states in e-e
 $N=Z$ nuclei are $t=0$ unpaired [or at most vibrational].
  This conclusion is in nice agreement with 
direct calculations of the ground-states in $N-Z=4$ nuclei. These
 isobaric analogue states are predicted systematically to be $pn$-unpaired by
 various types of models~\cite{[Eng96],[Eng97],[Rop00]}. The results
 of the $\Delta E_{T=2}$ excitation energy calculations are summerized in
 Fig.~\ref{all}c .
 Provided the simplicity of the model the agreement is rather amazing.
 Let us also bring the reader's attention to certain details. For example the
 shell-structure, reflecting proportionality of the 
 $sp$ iso-inertia to the $sp$ energy
 splitting at the Fermi surface, is quite clearly seen. Simoultanously,
 let us also note that these shell-structure effects are smeared out quite
 substantially by $t=0$ correlations as
 compared to calculations which include only $t=1$ pairing, see
 Fig~\ref{t2t1}. For more details showing the basic features of these
 calculations we refere the reader to~\cite{[Sat00a]}.

 \subsubsection{The T=1 states in even-even nuclei}

  Let us now turn our attention 
to $T=1$ states in e-e nuclei. Guided by the $sp-$model we
  will treat these states as the lowest $2qp$ configurations.
   We  block the two lowest quasiparticles
  in the original Woods-Saxon basis by exchanging:
  \be
  \left( \ba{c} \mbox{\boldmath{U}}_K \\  \mbox{\boldmath{V}}_K \ea \right)
  \quad \longrightarrow  \quad
  \left( \ba{c} \mbox{\boldmath{V}}_K^* \\ \mbox{\boldmath{U}}_K^* \ea \right)
  \ee
  for [$K=1,2$] while solving the HFB(LN) equations.
  Since at iso-frequency $\omega_\tau = 0$ the lowest $2qp$ state carries
  zero iso-alignment we impose the iso-cranking
condition  $T_x=1$, as shown in
  Fig.~\ref{all}a

  At $\omega_\tau = 0$ the $2qp$ state mixes both $t=0$ and $t=1$
  pairing phases [both in BCS as well as BCSLN approximations].
  However, different to the e-e iso-ground-state-band, subsequent iso-cranking
  of the $2qp$ state does not affect strongly isoscalar but quenches isovector
  pairing correlations. The $pp-$ and $nn-$pairing correlations disappear
  completely, exactly when the iso-alignment reaches unity. 
  At this point the system becomes 
  trapped and the state of the nucleus does not change till the very
  high frequency -- high enough to destroy isoscalar correlations. Note that,
  Fig.~\ref{all}c, the excitation energies of these states $\Delta E_{T=1}$
  again agree surprisingly well with empirical data.

\begin{figure}
\begin{center}
\leavevmode
\epsfysize=10.0cm
\epsfbox{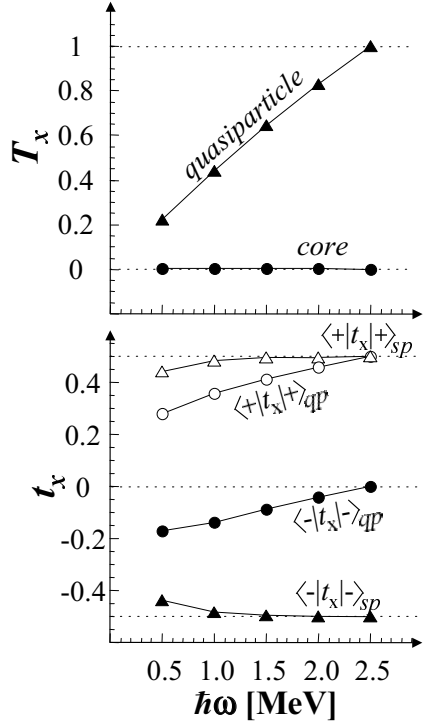}
\end{center}
\caption[]{Contributions of the  core and the  quartet of canonical 
quasiparticles to the total iso-alignment $T_x$ for the case of an e-e 
nucleus with isospin $T=1$. 
The  lower part illustrates the contributions of 
the individual canonical $sp$ and $qp$ states to the total iso-alignment
as a function of $\omega_\tau$. See text for details.
}
\label{figx}
\end{figure}

  Certain clues explaining this seemingly counterintuitive picture
  can be obtained by going to the canonical basis in which the
 the density matrix is diagonal.
 One observes that:
   \begin{enumerate}
   \item  For mixed-phases solution [below $T_x =1$]
   the blocking of the original Woods-Saxon states is not equivalent to 
   the blocking of  the canonical states i.e. all canonical 
   quasiparticles have fractional
   occupation numbers.
   \item With increasing frequency the occupation probability of the two lowest
   canonical quasiparticles increases gradually and, at $T_x=1$,
  reaches exactly unity. In this state one pair decouples from the
  purely $t=0$ $pn-$paired core and the isospins are aligned 
  along cranking axis.
   \item  At high frequency these canonical quasiparticles are built either on
   symmetric $|+\rangle \sim |n\rangle + |p\rangle$ or asymmetric
   $|-\rangle\sim |n\rangle - |p\rangle$ combinations of the initial
   Woods-Saxon $sp$ states.
  \end{enumerate}
  These observations are illustrated in Fig.~\ref{figx}. 
  The upper pannel of the figure clearly shows that the e-e core 
  is inert. Indeed, the entire iso-alignment can be traced back to
  four canonical quasiparticles emerging from the degenerate 
   [at $\omega_\tau=0$] quartet of $sp$ states. At low frequencies 
  all quasiparticles forming this quartet contribute to the iso-alignment
  i.e. they all have fractional occupations, see open and black dots in
  the lower part of Fig.~\ref{figx}. 
  With increasing $\omega_\tau$, the  occupation probability of 
  the energetically favored pair of canonical quasiparticles 
  increase until they reach unity while the energetically unfavored pair 
  becomes empty. Simoultanously, the  
  iso-alignments of the $sp$ canonical states building these 
  quasiparticles approach exactly $\pm 1/2$ 
  i.e. they are built on $\sim |n\rangle \pm |p\rangle$ combinations 
  of the original basis, see the discussion in sect.~\ref{spmodel}.
  Clearly, the  process of iso-cranking restores 
  the isospin symmetry [$T_x$=1] of the blocked $2qp$ state and, in 
  fact, the symmetry of the whole system which is composed by 
  a $t=0$ paired o--o core and a single, 
properly coupled pair of quasiparticles.

  Let us further note that the $t=1$ and $t=0$ pairing correspond to an
  entirely different scattering processes. This becomes evident
  after transforming the pair operators to
  $|\pm\rangle$ basis:
  \be
  P^\dagger_{n\bar n} + P^\dagger_{p\bar p}
  \longrightarrow a^\dagger_+ {\bar a}^\dagger_{+} + a^\dagger_- {\bar
  a}^\dagger_{-} \quad \mbox{and} \quad P^\dagger_{t=0} \longrightarrow
  a^\dagger_+ {\bar a}^\dagger_{-} - a^\dagger_- {\bar a}^\dagger_{+}.
  \ee
  Since with increasing $\omega_\tau$ the 
canonical basis approach $|\pm\rangle$
  states, the contribution to the $t=1$ pair field from the lowest $qp$ states
  is $\quad \propto U_+V_+^*\rightarrow 0$. In other words,
  with increasing $\omega_\tau$ blocking is more and more effective in the 
$t=1$  channel. Simoultanously, contributions of the blocked $qp$ state
  to the $t=0$ pair field are:
  $\quad \propto U_-V_+^* + U_+V_-^* \rightarrow 1 $. It means that
  $t=0$ pairing is rather stable with increasing $\omega_\tau$.
  Only at very high $\omega_\tau$, when the cranking energy
  will overcome the coherence of $t=0$ pairing, the phase transition
  to the $t=1$ paired system will take place 
  [see discussion in subsect.~\ref{st2}].

 \subsubsection{The T=0 and T=1 states in odd-odd nuclei}

  Let us finally discuss briefly $N=Z$ o-o nuclei.
  The lowest $T=0$ state ($T=0$ ground state) in o-o nuclei cannot be
  treated on the same footing as the ground state of the neighbouring
  $N=Z$ e-e nuclei, see also \cite{[Fra99]}. 
  The major argument stems from the simple
  empirical fact
  that all $T=0$ {\it ground-states\/} in o-o nuclei have $I\ne 0$. Therefore,
  their treatment requires the 
  explicit breaking of time-reversal symmetry as it was
  already discussed within the $sp-$model. 
  This can be achieved only by treating
  this state as a $2qp$ configuration.
  In contrast, the lowest $T=1$ states in o-o nuclei are expected to be
  seniority-zero states. The basic argument comes from the isobaric
  symmetry.  These $I=0$
  states form a triplet of isobaric analogue states together with the ground
  states of e-e $N-Z=\pm 2$ nuclei. Hence, they all should have similar
  structure.
  A simple calculation scheme [see Fig.~\ref{all}b] emerges from these
  considerations: ({\it i\/}) The $T=0$ states should be treated as
  $2qp$ configurations i.e. seniority-two states typical for all o-o
  nuclei. ({\it ii\/}) The $T=1$ states
  are HFB(LN) e-e like vaccua [false vaccua] excited in isospace
  by means of iso-cranking. Therefore they are 
  seniority-zero states similar to e-e nuclei.

  The relative excitation energy $\Delta E = E_{T=1} - E_{T=0}$ resulting
  from our calculations is shown in Fig.~\ref{all}c.
  Since $2qp$ excitations are
  almost as costly  in energy as iso-cranking to $T=1$, therefore, it
  is not surprising that both states stay nearly degenerate.
  What surprises is that the model predicts not only the 
near-degeneracy but also
  certain details like the inversion of $T=1$ and $T=0$ excitations
  taking place around $f_{7/2}$ ($A\sim 40$) sub-shell nuclei.
  Let us mention that in our approach the isoscalar and isovector
  phases are present both in $T=1$ and $T=0$ states. It is therefore evident
  that the near-degeneracy of $T=1$ and $T=0$ states cannot be used as 
an argument
  to rule out isoscalar pairing as it was done by the
  Berkeley group~\cite{[Mac00s]}. The key lays in
  the understanding of the 
  underlying structure, which can be achieved by means of
a microscopic model only.

\section{Summary}

We have shown that {\it mean-field\/} based models which
incorporate both $t=0$ and $t=1$ pairing correlations and,
at least in approximate manner,  
number- and isospin projection are in principle 
capable to treat consistently $N\sim Z$ nuclei.
The number projection is treated here at the level of
the Lipkin-Nogami approximation~\cite{[Pra73],[Sat00]}
while isospin is restored using the isospin cranking formalism
~\cite{[Che78],[Sat00a]}.  
Within the model the e-e vacuum
is a mixed $t=0$ and $t=1$ phase 
state due to the spontaneus symmetry breaking introduced 
by number projection~\cite{[Sat97],[Sat00]}. The $T=2$
state in e-e nuclei is an e-e vacuum
calculated at the iso-cranking frequency $\omega\tau$ corresponding to
 $T_x = \sqrt6$ while $T=1$ state in
e-e nuclei is described as a 2qp state 
at which  $T_x = 1$. The $T=0$ states in o-o nuclei 
are 2qp excitations while $T=1$ states in o-o nuclei
are e-e-like vaccua [{\it false vaccua\/}] computed at
$\omega\tau$ corresponding  
to $T_x=\sqrt2$. In all cases $t=0$ superfluidity 
plays a crucial role in restoring the correct iso-MoI and
in turn, the exciatation energies $\Delta E_T$. 
In fact, thanks to the simplicity of the iso-cranking 
approximation, both the
role and response of the $t=0$ phase against iso-rotations can
be very simply and intuitively understood by a number of beautifull 
analogies to  well studied phenomena of high-spin physics.

\bigskip

  This work was supported in part
 the Polish Committee for Scientific Research (KBN),
 and by the G{\"o}ran Gustafsson
 Foundation.


\end{document}